\documentclass[11pt]{article}

\usepackage{amsfonts,amssymb,chet,dsfont,graphicx,mathrsfs,pgfplots,tikz}

\usetikzlibrary{external}
\newcommand{\braket}[2]{\langle#1|#2\rangle}

\DeclareMathOperator{\sinc}{sinc}

\title{Photon-Dark Photon Conversions in Extreme Background Electromagnetic Fields}

\author{Jean-Fran\c{c}ois Fortin$^{\ast,}$\email{jean-francois.fortin@phy.ulaval.ca} and Kuver Sinha$^{\dagger,}$\email{kuver.sinha@ou.edu}}

\affiliation{
$^\ast$D\'epartement de Physique, de G\'enie Physique et d'Optique,\\Universit\'e Laval, Qu\'ebec, QC G1V 0A6, Canada\\
$^\dagger$Department of Physics and Astronomy, University of Oklahoma, Norman, OK 73019, USA
}

\abstract{%
The mixing of photons with light pseudoscalars in the presence of external electromagnetic fields has been used extensively to search for axion-like-particles.  A similar effect for dark photon propagating states is usually not considered due to the Landau-Yang theorem.  We point out that mixing between photon and dark photon propagating states in background electromagnetic fields can indeed occur, in non-linear QED, through a four-photon vertex by integrating out the electron box diagram.  Starting from the Schwinger Lagrangian, we derive the equations of motion for dark photons interacting with the Standard Model photon through gauge kinetic terms.  We provide expressions for the perpendicular and parallel refractive indices in series expansions in the critical field strength, valid both in the strong and weak background field limits.  We then consider mixing between the photon-dark photon propagating system in the presence of pure electric and magnetic background fields, and work out the probability of conversion when the background fields are homogeneous.  We indicate outlines of the calculation in the inhomogeneous case, and finally express our results in the active-sterile basis, where we find that the mixing induced by background fields can lead to corrections to the tree-level mixing in the zero field limit that is usually considered to probe such systems.  Our results may find applications for probing photon-dark photon conversions in the vicinity of neutron stars and in table-top petawatt laser experiments.
}

\date{April 2019} 

\begin{document}

\maketitle

\toc


\section{Introduction}\label{SIntro}

The classic paper by Raffelt and Stodolsky \cite{Raffelt:1987im} laid much of the theoretical foundations for describing the mixing of the photon with other particles $X$, in the presence of an external electromagnetic field.  The particle $X$, presumably belonging to some extension of the Standard Model (SM), is required to have a two-photon vertex---a photon will then produce $X$ as it enters a region with a background electromagnetic field.  Furthermore, if $X$ is light or massless, it will mix with the photon and a coherent superposition of the two states will be produced.

The crucial requirement in all of the above is that $X$ must have a two-photon vertex.  This immediately restricts $X$ to have either spin zero or spin two.  The case of spin zero has in particular been extensively studied, with $X$ being an axion-like-particle $a$, and the relevant vertex being given by $\mathcal{L}\supset g_{a\gamma\gamma}aF_{\mu\nu}\tilde{F}^{\mu\nu}$ \cite{Peccei:1977hh,Peccei:1977ur,Wilczek:1977pj,Weinberg:1977ma}.  These ideas form the basis of the helioscope \cite{Anastassopoulos:2017ftl} and light-shining-through-wall \cite{Redondo:2010dp} classes of experiments, and have also been used to study the conversion of axions to photons and vice versa in galactic and neutron star magnetic fields \cite{Fortin:2018ehg,Fortin:2018aom}.  We refer to \cite{Graham:2015ouw,Marsh:2015xka} for recent reviews.

A similar treatment for dark photons---spin one particles interacting with the photon through the gauge kinetic term---is forbidden by the Landau-Yang theorem.  However, one can entertain the possibility of photon-dark photon conversions through effective four-photon vertices in non-linear QED \cite{Heisenberg:1935qt,Schwinger:1951nm,Adler:1971wn}.  For example, the electron box diagram can be integrated out at energies below the electron mass to obtain an effective four-photon vertex.  Two of the external states can be taken to be background fields, and the remaining two will constitute a coherent photon-dark photon superposition.

We hasten to clarify that there is a trivial mixture between the SM  and dark photon coming from the mixing parameter $\varepsilon$ typically introduced in the dark photon Lagrangian.  Since SM particles couple to a mixture of the SM and dark photon, the physically and observationally relevant transition is that between the active state emitted by a source and the active state absorbed by the measuring apparatus.  Conversely, the attenuation of a signal is due to the transition from the active state to the sterile state.  In this sense, there is thus a trivial ``mixing'' between the SM and dark photons in the tree-level Lagrangian, in the active-sterile basis.  Our main interest in this paper, in contrast, is in the conversion and mixing between \textit{propagating} photon and dark photon states.  This mixing of propagating states when background fields are present, and which we explicitly calculate, modifies and corrects the tree-level mixing of the active-sterile basis.  It is to find these corrections that is the main point of our work.

Our purpose in this paper is to lay the theoretical groundwork for photon-dark photon mixing in non-linear QED.  Starting from the Schwinger Lagrangian, we derive the equations of motion for dark photons interacting with the SM photon through the gauge kinetic term.  We provide expressions for the perpendicular and parallel refractive indices in series expansions in the critical field strength, valid both in the strong and weak background field limits.  We then consider mixing between the photon-dark photon in the presence of pure electric and magnetic background fields, and work out the probability of conversion when the background fields are homogeneous in the physical case of active state-sterile state conversion.  The final results are encapsulated in \eqref{EqnP} and \eqref{EqnPinhom} of our paper, for the cases where the background fields are homogeneous and inhomogeneous, respectively.

Non-linear QED is only relevant if the background insertions acquire values close to the quantum critical field strength $B_c=m_e^2/e=4.414\times10^{13}\,\text{G}$.  Our calculations are thus expected to make  modifications of the tree-level mixing between active and sterile states only in these rather extreme situations.  However, there is growing data in precisely such extreme situations with quantum-critical electromagnetic fields.  Our results may thus be applied to photon-dark photon conversions in magnetars and petawatt lasers, where the background fields can be near quantum critical strength.

We further note that the standard scenarios where active-sterile conversion occurs---light shining through walls, helioscopes, \textit{etc.}---have been studied in the weak magnetic field limit in \cite{Masso:2006gc,Ahlers:2007rd,Ahlers:2007qf,An:2013yfc}, in models with extra hidden sector fermions.  In our work, we generalize these calculations to find the mixing between propagating states in \textit{all}, and in particular \textit{strong}, background fields, and compute the resulting modification of the mixing between active and sterile states.  Moreover, we do not introduce any extra hidden sector fields apart from the dark photon.  Thus, the effects that we calculate are unavoidable in models of dark photons propagating in background fields.

In the case where the background insertion is a magnetic field, the natural target for our investigations is a neutron star, in particular a magnetar.  Magnetars are neutron stars characterized by extremely strong magnetic fields close to or sometimes exceeding the critical value \cite{Turolla:2015mwa,Beloborodov:2016mmx,Kaspi:2017fwg}, and have been recently studied by the authors in the context of axion to X-ray conversion \cite{Fortin:2018ehg,Fortin:2018aom}.  Na\"ively, one expects interesting modulation of the magnetar photon spectrum due to photon-dark photon mixing in the vicinity of its surface where the magnetic field is strongest.

In the case where the background insertion is an electric field, the relevant systems are the upcoming 10 PW-class optical laser systems.  We note that for these lasers, the intensities $I\sim10^{23}\,\text{W}/\text{cm}^2$ are still lower than the quantum critical value $I_c=E^2_c\sim10^{29}\,\text{W}/\text{cm}^2$ corresponding to the electric field strength $E_c=m_e^2/e=1.3\times10^{18}\,\text{V/m}$.  Non-linear QED effects are nevertheless an important target for these systems \cite{Bragin:2017yau,Hill:2017uxy} and photon-dark photon conversions may be relevant.

Our paper is structured as follows.  In Section \ref{SDark}, we introduce the photon-dark photon model, and derive the equations of motion for the fields from the Schwinger Lagrangian.  In Section \ref{SWave}, we consider the propagation of the fields in pure external electric or magnetic backgrounds.  Expressions for the parallel and perpendicular refractive indices are obtained, and the mixing matrix is derived.  We also show the modified expressions with the inclusion of plasma effects.  In Section \ref{SConsequences}, we show the probability of conversion in homogeneous background fields and discuss briefly the inhomogeneous case.  We end with our conclusions in Section \ref{SConclusion}.
 

\section{Dark Photons}\label{SDark}

In this section the basic formalism describing dark photons and their couplings to the SM is introduced.  The effective Lagrangian at energies below the electron mass is presented and the equations of motion for the physical photon and dark photon are obtained.


\subsection{Photon-Dark Photon Lagrangian}

In extensions of the SM with massive dark photons \cite{Fayet:1980ad,Fayet:1990wx}, a dark sector $\mathscr{L}_\text{D}$ interacts with the SM $\mathscr{L}_\text{SM}$ solely through kinetic mixing $\mathscr{L}_{\text{SM}\otimes\text{D}}$ between the SM $U(1)_Y$ hypercharge gauge boson and the dark photon.  The kinetic mixing is generated through loops by integrating out heavy particles charged under both gauge groups \cite{Holdom:1985ag}, leading to
\eqn{\mathscr{L}=\mathscr{L}_\text{SM}+\mathscr{L}_\text{D}+\mathscr{L}_{\text{SM}\otimes\text{D}},\qquad\text{where}\qquad\mathscr{L}
_{\text{SM}\otimes\text{D}}=\frac{\varepsilon_Y}{2}F_{\mu\nu}^\text{SM}F_\text{D}^{\mu\nu}.}[EqnLhigh]
The corresponding SM and dark field strengths are $F_{\mu\nu}^\text{SM}=\partial_\mu B_\nu-\partial_\nu B_\mu$ and $F_\text{D}^{\mu\nu}=\partial^\mu A_\text{D}^\nu-\partial^\nu A_\text{D}^\mu$ respectively, where $B^\mu$ is the $U(1)_Y$ gauge boson and $A_\text{D}^\mu$ is the $U(1)_\text{D}$ gauge boson, \textit{i.e.} the dark photon.  Thus, at energies above the electroweak scale but below the heavy charged particle masses, the strength of the kinetic mixing between the SM hypercharge gauge group $U(1)_Y$ and the dark Abelian gauge group $U(1)_\text{D}$ generated by integrating out the heavy charged particles is parametrized by the dimensionless parameter $\varepsilon_Y$, which is naturally small $\varepsilon_Y\ll1$.

At energies below the electroweak scale but above the electron mass, the electroweak gauge group is broken and the mixing now occurs between the electromagnetic and dark gauge groups instead, with the mixing parameter $\varepsilon=\varepsilon_Y\cos\theta_W$ where $\theta_W$ is the weak angle.  The two gauge bosons can then be rotated into each other such that the resulting gauge bosons have canonically-normalized kinetic terms, with the new gauge bosons representing the physical photon and dark photon respectively.  After this transformation, the SM fields become millicharged under the physical dark gauge group \cite{Cassel:2009pu,Hook:2010tw}, leading to
\eqn{\mathscr{L}_{\text{SM}\otimes\text{D}}=-\varepsilon eJ_\mu^\text{SM}A_\text{D}^\mu,}[EqnLSMD]
where $J_\mu^\text{SM}$ is the SM electromagnetic current.

At energies slightly above the lightest SM charged particle mass, \textit{i.e.} the electron mass $m_e$, the dark photon couples to the SM only through the electron.  Hence $J_\mu^\text{SM}=\bar{\psi}\gamma_\mu\psi$ and the effective Lagrangian \eqref{EqnLhigh} with \eqref{EqnLSMD} becomes
\eqn{\mathscr{L}=-\frac{1}{4}F_{\mu\nu}^\text{SM}F_\text{SM}^{\mu\nu}+\bar{\psi}\gamma^\mu(i\partial_\mu-eA_\mu^\text{SM})\psi-m_e\bar{\psi}\psi-\frac{1}{4}F_{\mu\nu}^\text{D}F_\text{D}^{\mu\nu}+\frac{1}{2}m_\text{D}^2A_\mu^\text{D}A_\text{D}^\mu-\varepsilon e\bar{\psi}\gamma^\mu\psi A_\mu^\text{D},}[EqnLlow]
where $m_\text{D}$ is the dark photon mass.  Therefore the electron current effectively couples to the gauge boson $A_\mu=A_\mu^\text{SM}+\varepsilon A_\mu^\text{D}$, which is the active state.  We refer to \cite{Alexander:2016aln} for a review of experimental searches and constraints on such scenarios.

At energies below the electron mass but above the dark photon mass, the electron is integrated out.  For the SM without dark photons, this leads to usual non-linear QED effects through the electron box diagram, as first computed by Schwinger \cite{Heisenberg:1935qt,Schwinger:1951nm,Adler:1971wn,Heyl:1997hr},
\eqna{
\mathscr{L}_\text{S}(A_\mu^\text{SM})&=\frac{\alpha}{2\pi}\int d\zeta\,\frac{e^{-\zeta}}{\zeta^3}\left[i\zeta^2\frac{\sqrt{K_\text{SM}}}{4}\right.\\
&\left.\phantom{=}\qquad\times\frac{\cos\left(\frac{\zeta}{B_c}\sqrt{-\frac{I_\text{SM}}{2}+i\frac{\sqrt{K_\text{SM}}}{2}}\right)+\cos\left(\frac{\zeta}{B_c}\sqrt{-\frac{I_\text{SM}}{2}-i\frac{\sqrt{K_\text{SM}}}{2}}\right)}{\cos\left(\frac{\zeta}{B_c}\sqrt{-\frac{I_\text{SM}}{2}+i\frac{\sqrt{K_\text{SM}}}{2}}\right)-\cos\left(\frac{\zeta}{B_c}\sqrt{-\frac{I_\text{SM}}{2}-i\frac{\sqrt{K_\text{SM}}}{2}}\right)}+|B_c|^2+\frac{\zeta^2}{6}I_\text{SM}\right],}[EqnSchwinger]
where
\eqn{I_\text{SM}=F_{\mu\nu}^\text{SM}F_\text{SM}^{\mu\nu}=2(\boldsymbol{B}_\text{SM}^2-\boldsymbol{E}_\text{SM}^2),\qquad K_\text{SM}=(F_{\mu\nu}^\text{SM}\tilde{F}_\text{SM}^{\mu\nu})^2=(-4\boldsymbol{E}_\text{SM}\cdot\boldsymbol{B}_\text{SM})^2,}[EqnIKSM]
with the dual SM field strength given by $\tilde{F}_\text{SM}^{\mu\nu}=\tfrac{1}{2}\varepsilon^{\mu\nu\lambda\rho}F_{\lambda\rho}^\text{SM}$, and $B_c=m_e^2/e=4.414\times10^{13}\,\text{G}=E_c=1.318\times10^{18}\,\text{V}/\text{m}$ the critical QED field strength.

For the SM with dark photons, \eqref{EqnLlow} implies that the non-linear Lagrangian generated by the electron box diagram corresponds to the Schwinger Lagrangian \eqref{EqnSchwinger} with $A_\mu^\text{SM}\to A_\mu=A_\mu^\text{SM}+\varepsilon A_\mu^\text{D}$, leading to
\eqn{\mathscr{L}=-\frac{1}{4}(I_\text{SM}+I_\text{D})+\frac{1}{2}m_\text{D}^2A_\mu^\text{D}A_\text{D}^\mu+\mathscr{L}_\text{S}(A_\mu),}[EqnLNL]
after proper renormalization.  The shift from $A_\mu^\text{SM}$ to $A_\mu=A_\mu^\text{SM}+\varepsilon A_\mu^\text{D}$ implies $F_{\mu\nu}^\text{SM}\to F_{\mu\nu}=F_{\mu\nu}^\text{SM}+\varepsilon F_{\mu\nu}^\text{D}$ and the corresponding shifts
\eqna{
I_\text{SM}&\to I=F_{\mu\nu}F^{\mu\nu}=I_\text{SM}+2\varepsilon I_{\text{SM}\otimes\text{D}}+\varepsilon^2I_\text{D},\\
K_\text{SM}&\to K=(F_{\mu\nu}\tilde{F}^{\mu\nu})^2=\left(\sqrt{K_\text{SM}}+2\varepsilon\sqrt{K_{\text{SM}\otimes\text{D}}}+\varepsilon^2\sqrt{K_\text{D}}\right)^2,
}[EqnIKp]
for \eqref{EqnIKSM} in the Schwinger Lagrangian \eqref{EqnSchwinger} appearing in \eqref{EqnLNL}.  Here
\eqna{
I_{\text{SM}\otimes\text{D}}&=F_{\mu\nu}^\text{SM}F_\text{D}^{\mu\nu}=2(\boldsymbol{B}_\text{SM}\cdot\boldsymbol{B}_\text{D}-\boldsymbol{E}_\text{SM}\cdot\boldsymbol{E}_\text{D}),\\
K_{\text{SM}\otimes\text{D}}&=(F_{\mu\nu}^\text{SM}\tilde{F}_\text{D}^{\mu\nu})^2=[-2(\boldsymbol{E}_\text{SM}\cdot\boldsymbol{B}_\text{D}+\boldsymbol{E}_\text{D}\cdot\boldsymbol{B}_\text{SM})]^2,
}[EqnIKD]
effectively mixing the physical photon and dark photon at the non-linear level.

Following \cite{Heyl:1997hr}, the Schwinger Lagrangian in \eqref{EqnLNL} can be Taylor-expanded for small $K$, leading to
\eqn{\mathscr{L}_\text{S}(A_\mu)\equiv\mathscr{L}_\text{S}(I,K)=\mathscr{L}_\text{S}(I,0)+K\left.\frac{\partial\mathscr{L}_\text{S}(I,K)}{\partial K}\right|_{K=0}+\cdots.}[EqnTaylor]
In terms of the dimensionless parameter $\xi=\sqrt{I/(2B_c^2)}$, the different terms in \eqref{EqnTaylor} can be expressed as
\eqn{\mathscr{L}_\text{S}(I,0)=\frac{\alpha}{2\pi}\frac{I}{2}X_0(1/\xi),\qquad\left.\frac{\partial\mathscr{L}_\text{S}(I,K)}{\partial K}\right|_{K=0}=-\frac{\alpha}{2\pi}\frac{1}{16I}X_1(1/\xi),}[EqnTerms]
where
\eqna{
X_0(x)&=4\int_0^{x/2-1}dy\,\ln[\Gamma(y+1)]+\frac{1}{3}\ln(1/x)+2\ln(4\pi)-4\left[\frac{1}{12}-\zeta^{(1)}(-1)\right]-\frac{5}{3}\ln(2)\\
&\phantom{=}\qquad-\left[\ln(4\pi)+1+\ln(1/x)\right]x+\left[\frac{3}{4}+\frac{1}{2}\ln(2/x)\right]x^2,\\
X_1(x)&=-2X_0(x)+xX_0^{(1)}(x)+\frac{2}{3}X_0^{(2)}(x)-\frac{2}{9}\frac{1}{x^2},
}[EqnX]
with superscripts in parenthesis denoting differentiation with respect to the appropriate argument, \textit{e.g.} $X_0^{(n)}(x)=d^nX_0(x)/dx^n$.

Therefore, the starting point for the analysis of photon-dark photon oscillations is the effective Lagrangian \eqref{EqnLNL} with the expansion \eqref{EqnTaylor} and its different contributions \eqref{EqnTerms},
\eqn{\mathscr{L}=-\frac{1}{4}(I_\text{SM}+I_\text{D})+\frac{1}{2}m_\text{D}^2A_\mu^\text{D}A_\text{D}^\mu+\frac{\alpha}{2\pi}\left[\frac{I}{2}X_0(1/\xi)-\frac{K}{16I}X_1(1/\xi)\right]+\cdots,}[EqnLosc]
the latter being given explicitly by \eqref{EqnX}.

Since $X_0(x)$ behaves as
\eqn{X_0(x)\sim\frac{1}{45x^2}-\frac{4}{315x^4}+\frac{8}{315x^6}-\frac{32}{297x^8}+\cdots,}
at large $x$, the effective Lagrangian \eqref{EqnLNL} in the weak-field limit is
\eqn{\mathscr{L}=-\frac{1}{4}(I_\text{SM}+I_\text{D})+\frac{1}{2}m_\text{D}^2A_\mu^\text{D}A_\text{D}^\mu+\frac{\alpha^2}{90m_e^4}\left(I^2+\frac{7K}{4}\right)+\cdots,}
in accord with the Euler-Heisenberg Lagrangian \cite{Heisenberg:1935qt} in the pure QED case.


\subsection{Equations of Motion}

From the following identities,
\eqn{\frac{\partial I}{\partial(\partial_\mu A_\nu^\text{SM})}=\frac{1}{\varepsilon}\frac{\partial I}{\partial(\partial_\mu A_\nu^\text{D})}=4F^{\mu\nu},\qquad\frac{\partial K}{\partial(\partial_\mu A_\nu^\text{SM})}=\frac{1}{\varepsilon}\frac{\partial K}{\partial(\partial_\mu A_\nu^\text{D})}=8\sqrt{K}\tilde{F}^{\mu\nu},}
and \eqref{EqnLosc}, the equations of motion for the physical photon and dark photon are
\eqn{\partial_\mu F_\text{SM}^{\mu\nu}=\frac{\alpha}{2\pi}\partial_\mu Z_1^{\mu\nu}(I,K)+\cdots,\qquad\partial_\mu F_\text{D}^{\mu\nu}+m_\text{D}^2A_\text{D}^\nu=\frac{\varepsilon\alpha}{2\pi}\partial_\mu Z_1^{\mu\nu}(I,K)+\cdots,}[EqnEOM]
with
\eqna{
Z_1^{\mu\nu}(I,K)&=\left[2X_0(1/\xi)-\frac{1}{\xi}X_0^{(1)}(1/\xi)+\frac{K}{16B_c^4}\frac{1}{\xi^4}X_1(1/\xi)+\frac{K}{32B_c^4}\frac{1}{\xi^5}X_1^{(1)}(1/\xi)\right]F^{\mu\nu}\\
&\phantom{=}\qquad-\frac{\sqrt{K}}{4B_c^2}\frac{1}{\xi^2}X_1(1/\xi)\tilde{F}^{\mu\nu}.
}[EqnZ]
The equations of motion \eqref{EqnEOM} can be simplified to
\eqn{\partial_\mu F^{\mu\nu}+\varepsilon m_\text{D}^2A_\text{D}^\nu=\frac{\alpha}{2\pi}(1+\varepsilon^2)\partial_\mu Z_1^{\mu\nu}(I,K)+\cdots,\qquad\varepsilon\partial_\mu F_\text{SM}^{\mu\nu}=\partial_\mu F_\text{D}^{\mu\nu}+m_\text{D}^2A_\text{D}^\nu,}[EqnEOMp]
and the derivative of \eqref{EqnZ} can be evaluated to
\eqna{
\partial_\mu Z_1^{\mu\nu}(I,K)&=\left[2X_0(1/\xi)-\frac{1}{\xi}X_0^{(1)}(1/\xi)+\frac{K}{16B_c^4}\frac{1}{\xi^4}X_1(1/\xi)+\frac{K}{32B_c^4}\frac{1}{\xi^5}X_1^{(1)}(1/\xi)\right]\partial_\mu F^{\mu\nu}\\
&\phantom{=}\qquad-\frac{1}{\xi^3}\left[X_0^{(1)}(1/\xi)-\frac{1}{\xi}X_0^{(2)}(1/\xi)+\frac{K}{4B_c^4}\frac{1}{\xi^3}X_1(1/\xi)+\frac{7K}{32B_c^4}\frac{1}{\xi^4}X_1^{(1)}(1/\xi)\right.\\
&\phantom{=}\qquad\left.+\frac{K}{32B_c^4}\frac{1}{\xi^5}X_1^{(2)}(1/\xi)\right]\frac{F_{\alpha\beta}F^{\mu\nu}}{2B_c^2}\partial_\mu F^{\alpha\beta}\\
&\phantom{=}\qquad+\frac{\sqrt{K}}{2B_c^2}\frac{1}{\xi^4}\left[X_1(1/\xi)+\frac{1}{2\xi}X_1^{(1)}(1/\xi)\right]\frac{\tilde{F}_{\alpha\beta}F^{\mu\nu}}{2B_c^2}\partial_\mu F^{\alpha\beta}\\
&\phantom{=}\qquad-\frac{\sqrt{K}}{4B_c^2}\frac{1}{\xi^2}X_1(1/\xi)\partial_\mu\tilde{F}^{\mu\nu}+\frac{\sqrt{K}}{4B_c^2}\frac{1}{\xi^4}\left[2X_1(1/\xi)+\frac{1}{\xi}X_1^{(1)}(1/\xi)\right]\frac{F_{\alpha\beta}\tilde{F}^{\mu\nu}}{2B_c^2}\partial_\mu F^{\alpha\beta}\\
&\phantom{=}\qquad-\frac{1}{\xi^2}X_1(1/\xi)\frac{\tilde{F}_{\alpha\beta}\tilde{F}^{\mu\nu}}{2B_c^2}\partial_\mu F^{\alpha\beta}.
}[EqndZ]
Expressing all the field strengths with extra derivatives as second-order derivatives on the gauge fields and the remaining field strengths as electric and magnetic fields, the equations of motion \eqref{EqnEOMp} become
\eqna{
\partial^2A^\nu+\varepsilon m_\text{D}^2A_\text{D}^\nu-\partial^\nu(\partial\cdot A)&=\frac{\alpha}{2\pi}(1+\varepsilon^2)\partial_\mu Z_1^{\mu\nu}(I,K)+\cdots,\\
\varepsilon\partial^2 A_\text{SM}^\nu-\varepsilon\partial^\nu(\partial\cdot A_\text{SM})&=(\partial^2+m_\text{D}^2)A_\text{D}^\nu-\partial^\nu(\partial\cdot A_\text{D}),
}[EqnEOMpp]
where all $F_{\alpha\beta}F^{\mu\nu}$, $\tilde{F}_{\alpha\beta}F^{\mu\nu}$, $F_{\alpha\beta}\tilde{F}^{\mu\nu}$ and $\tilde{F}_{\alpha\beta}\tilde{F}^{\mu\nu}$ in \eqref{EqndZ} are functions of the external electric and magnetic fields only.  It is easy to notice that $\partial_\nu$ applied on \eqref{EqnEOMpp} leads to the condition $\partial\cdot A_\text{D}=0$ as long as $m_\text{D}\neq0$.  Moreover, standard gauge fixing can be done with the help of $\partial\cdot A_\text{SM}=0$ supplemented with $A_t^\text{SM}=0$.

For the purpose of wave propagation, it remains to properly normalize the spatial derivative terms in the equations of motion \eqref{EqnEOMpp} to determine the refractive indices.  In the small $K$ limit, this can be done in two relevant cases, \textit{i.e.} for pure external magnetic field or pure external electric field, for which $K=0$.  In these two physically-relevant cases, \eqref{EqndZ} reduces to
\eqna{
\partial_\mu Z_1^{\mu\nu}(I,K)&=\left[2X_0(1/\xi)-\frac{1}{\xi}X_0^{(1)}(1/\xi)\right]\partial_\mu F^{\mu\nu}\\
&\phantom{=}\qquad-\left[\frac{1}{\xi}X_0^{(1)}(1/\xi)-\frac{1}{\xi^2}X_0^{(2)}(1/\xi)\right]\frac{F_{\alpha\beta}F^{\mu\nu}}{I}\partial_\mu F^{\alpha\beta}-X_1(1/\xi)\frac{\tilde{F}_{\alpha\beta}\tilde{F}^{\mu\nu}}{I}\partial_\mu F^{\alpha\beta}.
}[EqndZK0]
It is now straightforward to obtain the equations of motion for propagation in the $z$ direction in these two cases from \eqref{EqnEOMpp} and \eqref{EqndZK0}.  The chosen convention for the parallel and perpendicular modes is the usual one where the parallel mode is the propagating mode with its electric field in the plane spanned by the external field and the direction of propagation while the perpendicular mode is the propagating mode with its electric field perpendicular to that plane.

Before proceeding, it is worth mentioning that for time-independent external fields, the time coordinate can be Fourier-transformed to the angular frequency $\omega$ such that $\partial_t\to-i\omega$.  In fact, both the SM photon and dark photon time components do not propagate [as can be seen from \eqref{EqnEOMpp}] and can be solved algebraically in terms of the transverse and (dark photon) longitudinal modes.

For the SM photon, the equation of motion for the longitudinal mode is trivially satisfied once the solution for the time components have been substituted, leading to two transverse propagating degrees of freedom as expected from gauge invariance.  Moreover, for relativistic dark photons, the dark photon mass is negligible and the dark photon longitudinal mode can be discarded.  Therefore, the dark photon mass can be understood as a contribution to the refractive indices of the dark photon transverse modes.

Finally, the remaining contributions to the refractive indices are $\alpha$-suppressed and thus small as long as the external fields are not too large.  Therefore the weak dispersion limit can be used which implies $\partial_z\to in\omega$.  Indeed, for external fields with spatial variations on much larger scales than the wavelength, the dispersion relation would be $k=n\omega$ with refractive index $n$.  But in the weak dispersion limit $|n-1|\ll1$, therefore the substitution $\partial_z\to i\omega$ is appropriate for all terms in the equations of motion that are already suppressed.


\section{Wave Propagation in a Pure External Field}\label{SWave}

This section discusses wave propagation in an external field taking into account non-linear effects from the vacuum contributions.  Wave propagation is first discussed without considering plasma effects, the latter being included subsequently.


\subsection{Vacuum Contributions without Plasma Effects}

In an external magnetic field with $\boldsymbol{E}_\text{SM}=\boldsymbol{E}_\text{D}=0$ or an external electric field with $\boldsymbol{B}_\text{SM}=\boldsymbol{B}_\text{D}=0$ such that $K=0$ and $I=2\boldsymbol{B}^2$ or $I=-2\boldsymbol{E}^2$,\footnote{Technically speaking, the only requirement for $K=0$ is that either $\boldsymbol{E}=0$ or $\boldsymbol{B}=0$.  Hence configurations where both visible and dark fields are non-vanishing but their properly-weighted combination vanishes are allowed.  Such configurations could occur in physical settings in the presence of extra matter fields charged under the dark gauge group.} the equations of motion \eqref{EqnEOMpp} for wave propagation in the $z$ direction to lowest non-trivial order in $\alpha$ and $\varepsilon$ are
\eqn{\left[\omega^2+\partial_z^2+\left(\begin{array}{cccc}Q_\perp&\varepsilon Q_\perp&0&0\\\varepsilon Q_\perp&-m_\text{D}^2&0&0\\0&0&Q_\parallel&\varepsilon Q_\parallel\\0&0&\varepsilon Q_\parallel&-m_\text{D}^2\end{array}\right)\right]\left(\begin{array}{c}A_\perp^\text{SM}\\A_\perp^\text{D}\\A_\parallel^\text{SM}\\A_\parallel^\text{D}\end{array}\right)=0,}[EqnEOMExt]
with
\eqn{Q_\perp^B=Q_\parallel^E=\frac{\alpha}{2\pi}\left[\frac{1}{\xi^2}X_0^{(2)}(1/\xi)-\frac{1}{\xi}X_0^{(1)}(1/\xi)\right]\omega^2\sin^2\theta,\qquad Q_\parallel^B=Q_\perp^E=-\frac{\alpha}{2\pi}X_1(1/\xi)\omega^2\sin^2\theta,}[EqnQ]
and $\xi=|\boldsymbol{B}|/B_c=b$ or $\xi=i|\boldsymbol{E}|/E_c=iy$ as dictated by \eqref{EqndZK0}.  Since $Q_i=2\omega^2(n_i-1)$ in the weak dispersion limit, the refractive indices are
\eqn{
\begin{gathered}
n_\perp^B=n_\parallel^E=1+\frac{\alpha}{4\pi}\left[\frac{1}{\xi^2}X_0^{(2)}(1/\xi)-\frac{1}{\xi}X_0^{(1)}(1/\xi)\right]\sin^2\theta,\\
n_\parallel^B=n_\perp^E=1-\frac{\alpha}{4\pi}X_1(1/\xi)\sin^2\theta,\qquad n_\text{D}=1-\frac{m_\text{D}^2}{2\omega^2},
\end{gathered}
}[Eqnn]
in agreement with \cite{Heyl:1997hr}.

For an external field with spatial variations on much larger scales than the wavelength, the system of second-order differential equations \eqref{EqnEOMExt} can be simplified with the help of $\omega^2+\partial_z^2=(\omega-i\partial_z)(\omega+i\partial_z)\to2\omega(\omega+i\partial_z)$ \cite{Raffelt:1987im}, leading to
\eqn{\left[\omega+i\partial_z+\left(\begin{array}{cccc}\Delta_\perp&\varepsilon\Delta_\perp&0&0\\\varepsilon\Delta_\perp&\Delta_\text{D}&0&0\\0&0&\Delta_\parallel&\varepsilon\Delta_\parallel\\0&0&\varepsilon\Delta_\parallel&\Delta_\text{D}\end{array}\right)\right]\left(\begin{array}{c}A_\perp^\text{SM}\\A_\perp^\text{D}\\A_\parallel^\text{SM}\\A_\parallel^\text{D}\end{array}\right)=0,}[EqnEOM1st]
with $\Delta_i=\omega(n_i-1)$, or more explicitly
\eqn{
\begin{gathered}
\Delta_\perp^B=\Delta_\parallel^E=\frac{\alpha}{4\pi}\left[\frac{1}{\xi^2}X_0^{(2)}(1/\xi)-\frac{1}{\xi}X_0^{(1)}(1/\xi)\right]\omega\sin^2\theta,\\
\Delta_\parallel^B=\Delta_\perp^E=-\frac{\alpha}{4\pi}X_1(1/\xi)\omega\sin^2\theta,\qquad\Delta_\text{D}=-\frac{m_D^2}{2\omega},
\end{gathered}
}[EqnDelta]
derived straightforwardly from \eqref{EqnQ} or \eqref{Eqnn}.

Following \cite{Heyl:1997hr}, \eqref{EqnDelta} can be simplified to
\eqn{
\begin{gathered}
\Delta_\perp^B=\frac{1}{2}q_\perp^B\omega\sin^2\theta,\qquad\Delta_\parallel^B=\frac{1}{2}q_\parallel^B\omega\sin^2\theta,\\
\Delta_\parallel^E=\frac{1}{2}q_\parallel^E\omega\sin^2\theta,\qquad\Delta_\perp^E=\frac{1}{2}q_\perp^E\omega\sin^2\theta,
\end{gathered}
}[EqnDeltaBp]
where the dimensionless quantities $q_\parallel$ and $q_\perp$ are functions of the pure external field given by \cite{Lai:2006af,Raffelt:1987im}
\eqn{
\begin{gathered}
q_\perp^B=\frac{4\alpha}{45\pi}b^2\hat{q}_\perp^B,\qquad\qquad\hat{q}_\perp^B=\frac{1}{1+(18/25)b^{5/4}+(4/15)b^2},\\
q_\parallel^B=\frac{7\alpha}{45\pi}b^2\hat{q}_\parallel^B,\qquad\qquad\hat{q}_\parallel^B=\frac{1+(5/4)b}{1+(133/100)b+(14/25)b^2},
\end{gathered}
}[EqnqB]
for pure external magnetic field and
\eqn{
\begin{array}{rl}
q_\perp^E=-\frac{7\alpha}{45\pi}y^2\hat{q}_\parallel^E,\qquad&\hat{q}_\perp^E=\frac{1+(81/56)y^2+(45/49)y^4}{1+(16/31)y^2+(35/24)y^4+(45/196)y^6}\\
&\qquad\qquad-i\frac{45}{14y^3}\left[\frac{2\pi}{3}\frac{ye^{-\pi/y}}{1-e^{-\pi/y}}-\ln(1-e^{-\pi/y})+\frac{2}{\pi}y\,\text{Li}_2(e^{-\pi/y})\right],\\
q_\parallel^E=-\frac{4\alpha}{45\pi}y^2\hat{q}_\perp^E,\qquad&\hat{q}_\parallel^E=\frac{1+(61/23)y^2-(40/23)y^4}{1-(35/58)y^2+(121/25)y^4+(32/69)y^6}\\
&\qquad\qquad-i\frac{45}{8y^4}\left[\pi\frac{e^{-\pi/y}}{1-e^{-\pi/y}}-y\ln(1-e^{-\pi/y})\right],
\end{array}
}[EqnqE]
for pure external electric field.  Here the real parts of $\hat{q}_\perp$ and $\hat{q}_\parallel$ are interpolating functions for the vacuum contributions to the refractive indices that are accurate to better than about 5\% almost everywhere while the imaginary parts of $\hat{q}_\perp$ and $\hat{q}_\parallel$ for a pure external electric field are exact.


\subsection{Vacuum Contributions with Plasma Effects}

The introduction of plasma effects is as simple as the introduction of vacuum contributions computed in the previous section.  Indeed, plasma effects can be seen as originating from the response of the charged particles present in the plasma to the passage of a visible photon.

In the present setting where the SM charged particles are millicharged under the dark gauge group, the passage of a visible photon leads to the usual response in the visible sector plus the same response in the dark sector, suppressed by the mixing parameter $\varepsilon$.

The same can be said of the passage of a dark photon, leading to the usual response in the visible sector, suppressed by the mixing parameter, plus the same response in the dark sector, suppressed by the mixing parameter square.  To first non-trivial order in $\varepsilon$, this last contribution vanishes.

Thus at lowest non-trivial order in $\alpha$ and $\varepsilon$, the simple system of first-order differential equations \eqref{EqnEOM1st} is modified to
\eqn{\left[\omega+i\partial_z+\left(\begin{array}{cccc}\Delta_\perp+\Delta_\perp^\text{pl}&\varepsilon(\Delta_\perp+\Delta_\perp^\text{pl})&\Delta_{\perp\to\parallel}^\text{pl}&\varepsilon\Delta_{\perp\to\parallel}^\text{pl}\\\varepsilon(\Delta_\perp+\Delta_\perp^\text{pl})&\Delta_\text{D}&\varepsilon\Delta_{\perp\to\parallel}^\text{pl}&0\\\Delta_{\parallel\to\perp}^\text{pl}&\varepsilon\Delta_{\parallel\to\perp}^\text{pl}&\Delta_\parallel+\Delta_\parallel^\text{pl}&\varepsilon(\Delta_\parallel+\Delta_\parallel^\text{pl})\\\varepsilon\Delta_{\parallel\to\perp}^\text{pl}&0&\varepsilon(\Delta_\parallel+\Delta_\parallel^\text{pl})&\Delta_\text{D}\end{array}\right)\right]\left(\begin{array}{c}A_\perp^\text{SM}\\A_\perp^\text{D}\\A_\parallel^\text{SM}\\A_\parallel^\text{D}\end{array}\right)=0,}[EqnEOM1stPl]
where the refractive indices are corrected due to the plasma effects.  Here, the plasma contributions neglecting the protons as in \cite{Lai:2006af} lead to
\eqn{
\begin{gathered}
\Delta_\perp^\text{pl}=-\frac{\omega_\text{pl}^2}{2\omega}\frac{\omega^2}{\omega^2-\omega_c^2},\qquad\Delta_\parallel^\text{pl}=-\frac{\omega_\text{pl}^2}{2\omega}\left(\sin^2\theta+\frac{\omega^2}{\omega^2-\omega_c^2}\cos^2\theta\right),\\
\Delta_{\perp\to\parallel}^\text{pl}=-\Delta_{\parallel\to\perp}^\text{pl}=-\frac{\omega_\text{pl}^2}{2\omega}\frac{i\omega\omega_c}{\omega^2-\omega_c^2}\cos\theta,
\end{gathered}
}[EqnPl]
where $\omega_\text{pl}=\sqrt{4\pi\alpha n_e/m_e}$ is the electron plasma frequency ($n_e$ is the electron density) and $\omega_c=\sqrt{\alpha}B/(m_ec)$ is the electron cyclotron frequency.

It is obvious that \eqref{EqnEOM1stPl} does not decouple nicely as \eqref{EqnEOM1st}.  Nevertheless, in the high-magnetization limit (where $\omega_c\gg\omega,\omega_\text{pl}$), the system simplifies greatly since
\eqn{\Delta_\perp^\text{pl}\to0,\qquad\Delta_\parallel^\text{pl}\to-\frac{\omega_\text{pl}^2}{2\omega}\sin^2\theta,\qquad\Delta_{\perp\to\parallel}^\text{pl}=-\Delta_{\parallel\to\perp}^\text{pl}\to0,}[EqnHighM]
leading to two decoupled systems of first-order differential equations given by
\eqn{
\begin{gathered}
\left[\omega+i\partial_z+\left(\begin{array}{cc}\Delta_\perp&\varepsilon\Delta_\perp\\\varepsilon\Delta_\perp&\Delta_\text{D}\end{array}\right)\right]\left(\begin{array}{c}A_\perp^\text{SM}\\A_\perp^\text{D}\end{array}\right)=0,\\
\left[\omega+i\partial_z+\left(\begin{array}{cc}\Delta_\parallel+\Delta_\parallel^\text{pl}&\varepsilon(\Delta_\parallel+\Delta_\parallel^\text{pl})\\\varepsilon(\Delta_\parallel+\Delta_\parallel^\text{pl})&\Delta_\text{D}\end{array}\right)\right]\left(\begin{array}{c}A_\parallel^\text{SM}\\A_\parallel^\text{D}\end{array}\right)=0.
\end{gathered}
}[EqnEOM1stPlHighM]
Therefore, in the high-magnetization limit \eqref{EqnHighM} and at lowest non-trivial order in $\alpha$ and $\varepsilon$, wave propagation for photon-dark photon system is described by \eqref{EqnEOM1stPlHighM}.  The system of first-order differential equations allows for independent oscillations between the perpendicular or the parallel modes of the photon-dark photon setup.  Moreover, plasma effects are negligible for oscillations of the perpendicular modes in the high-magnetization limit, as expected.


\section{Consequences}\label{SConsequences}

In this section the consequences of the results obtained previously (in the high-magnetization limit and at first non-trivial order in $\alpha$ and $\varepsilon$ for simplicity) are investigated in general terms following the work of \cite{Raffelt:1987im}.  The focus here is on the conversion probability, although the full analysis of \cite{Raffelt:1987im} can be straightforwardly redone for photon-dark photon oscillations.

Propagation in a homogeneous external field and in an inhomogeneous external field are discussed in turn.  For both the perpendicular modes and the parallel modes, the results are given in terms of the following system of first-order differential equations,
\eqn{\left[\omega+i\partial_z+\left(\begin{array}{cc}\Delta&\varepsilon\Delta\\\varepsilon\Delta&\Delta_\text{D}\end{array}\right)\right]\left(\begin{array}{c}A^\text{SM}\\A^\text{D}\end{array}\right)=0,}[EqnEOMA]
with proper substitutions, due to the symmetry between the system of first-order differential equations for the perpendicular modes and the parallel modes in the high-magnetization limit and at lowest non-trivial order in $\alpha$ and $\varepsilon$.


\subsection{Propagating versus Active/Sterile Bases}

Before proceeding, it is necessary to distinguish between the propagating (physical) states and the the active/sterile states that interact straightforwardly with SM matter.

Starting from the Lagrangian in the mass diagonal basis \eqref{EqnLlow}, we have provided results for propagating photon states $A^\text{SM}$ and propagating dark photon states $A^\text{D}$.  On the other hand, the state that is emitted or absorbed by a source (\textit{i.e.}, a charged matter particle belonging to the SM) is an active state denoted by $A^a$ while the associated orthogonal state, the sterile state denoted by $A^s$, does not interact with SM particles and is thus neither emitted by a source nor picked up by an observational instrument.  These states are given in terms of the propagating states by
\eqn{A^a\equiv\frac{A^\text{SM}+\varepsilon A^\text{D}}{\sqrt{1+\varepsilon^2}},\qquad\qquad A^s\equiv\frac{A^\text{D}-\varepsilon A^\text{SM}}{\sqrt{1+\varepsilon^2}}.}[EqAS]
We note that the longitudinal polarization of the dark photon is also an active state that interacts with SM charged particles at $\mathcal{O}(\varepsilon)$ \cite{Graham:2014sha}.  Although this longitudinal state can be important, as explained above it will be discarded here and all fields will be understood to be transverse.  The expressions for the active and sterile states \eqref{EqAS} can be obtained by starting from the interacting Lagrangian instead of the mass-diagonal Lagrangian \eqref{EqnLlow}.

For future convenience, we can define the transformation matrix for changing bases
\eqn{\left(\begin{array}{c}A^a\\A^s\end{array}\right)=\mathcal{R}\left(\begin{array}{c}A^\text{SM}\\A^\text{D}\end{array}\right),}[EqPtoAS]
where
\eqn{\mathcal{R}\equiv\frac{1}{\sqrt{1+\varepsilon^2}}\left(\begin{array}{cc}1&\varepsilon\\-\varepsilon&1\end{array}\right)\approx\left(\begin{array}{cc}1&\varepsilon\\-\varepsilon&1\end{array}\right).}[EqR]
For consistency, in the last part of \eqref{EqR} only the first non-trivial order in $\varepsilon$ is kept.  With \eqref{EqPtoAS} and \eqref{EqR}, it will be straightforward to compute the conversion probability in the proper basis.


\subsection{Homogeneous External Field}

For a homogeneous external field, the system of differential equations \eqref{EqnEOMA} can be easily Fourier-transformed with the replacement $\partial_z\to ik=in\omega$ and the appropriate refractive index.

Rotating the fields such as
\eqn{\left(\begin{array}{c}A^{\text{SM}'}\\A^{\text{D}'}\end{array}\right)=\left(\begin{array}{cc}\cos\vartheta&\sin\vartheta\\-\sin\vartheta&\cos\vartheta\end{array}\right)\left(\begin{array}{c}A^\text{SM}\\A^\text{D}\end{array}\right),}[EqnAp]
with the mixing angle $\vartheta$ satisfying
\eqn{\frac{1}{2}\tan(2\vartheta)=\frac{\varepsilon\Delta}{\Delta-\Delta_\text{D}},}[EqnMixing]
the system of propagation equations \eqref{EqnEOMA} become diagonal in terms of the rotated fields \eqref{EqnAp}.  Indeed, it corresponds to
\eqn{\left[\omega+i\partial_z+\left(\begin{array}{cc}\Delta_\text{SM}'&0\\0&\Delta_\text{D}'\end{array}\right)\right]\left(\begin{array}{c}A^{\text{SM}'}\\A^{\text{D}'}\end{array}\right)=0,}[EqnEOMAp]
where the prime quantities are explicitly given by
\eqn{\Delta_\text{SM}'=\frac{\Delta+\Delta_\text{D}}{2}+\frac{\Delta-\Delta_\text{D}}{2\cos(2\vartheta)},\qquad\Delta_\text{D}'=\frac{\Delta+\Delta_\text{D}}{2}-\frac{\Delta-\Delta_\text{D}}{2\cos(2\vartheta)}.}[EqnDeltap]
Hence, using \eqref{EqnMixing}, \eqref{EqnEOMAp} and \eqref{EqnDeltap} the rotated fields propagate as
\eqn{\left(\begin{array}{c}A^{\text{SM}'}\\A^{\text{D}'}\end{array}\right)(z)=e^{i\omega(z-t)}\left(\begin{array}{cc}e^{i\Delta_\text{SM}'z}&0\\0&e^{i\Delta_\text{D}'z}\end{array}\right)\left(\begin{array}{c}A^{\text{SM}'}\\A^{\text{D}'}\end{array}\right)(0),}
while the photon and dark photon states propagate following
\eqn{\left(\begin{array}{c}A^\text{SM}\\A^\text{D}\end{array}\right)(z)=e^{i\omega(z-t)}\mathcal{M}(z)\left(\begin{array}{c}A^\text{SM}\\A^\text{D}\end{array}\right)(0),}[EqnEOMhom]
where
\eqn{\mathcal{M}(z)=\left(\begin{array}{cc}\cos\vartheta&-\sin\vartheta\\\sin\vartheta&\cos\vartheta\end{array}\right)\left(\begin{array}{cc}e^{i\Delta_\text{SM}'z}&0\\0&e^{i\Delta_\text{D}'z}\end{array}\right)\left(\begin{array}{cc}\cos\vartheta&\sin\vartheta\\-\sin\vartheta&\cos\vartheta\end{array}\right).}[EqnM]

Clearly, the conversion probability in the proper basis is obtained from \eqref{EqPtoAS}, \eqref{EqR}, \eqref{EqnEOMhom} and \eqref{EqnM} as
\eqn{P_{\gamma_a\to\gamma_s}(z)=|\braket{A^s(z)}{A^a(0)}|^2=|(\mathcal{R}\mathcal{M}(z)\mathcal{R}^T)_{21}|^2,}
which is given explicitly by
\eqn{P_{\gamma_a\to\gamma_s}(z)=\left[1-\frac{\varepsilon}{\frac{1}{2}\tan(2\vartheta)}\right]^2\sin^2(2\vartheta)\sin^2\left[\frac{(\Delta-\Delta_\text{D})z}{2\cos(2\vartheta)}\right]=\frac{\Delta_\text{D}^2}{\Delta^2}\sin^2(2\vartheta)\sin^2\left[\frac{(\Delta-\Delta_\text{D})z}{2\cos(2\vartheta)}\right],}[EqnPH]
in the case of a pure external magnetic field after using \eqref{EqnMixing}.  The pure external electric field case is left to the reader, although it is clear that the conversion probability in that case is damped as a function of $z$ due to the imaginary contribution to the refractive indices as in \eqref{EqnqE} (contrary to the pure external magnetic field case for which the refractive indices \eqref{EqnqB} are real).

Using \eqref{EqnMixing} again in \eqref{EqnPH} gives
\eqn{P_{\gamma_a\to\gamma_s}(z)=\varepsilon^2\Delta_\text{D}^2z^2\sinc^2\left[\frac{(\Delta-\Delta_\text{D})z}{2}\right],}[EqnP]
to lowest non-trivial order in $\varepsilon$.  Here $\sinc(x)=\sin(x)/x$ is the cardinal sine function.  Hence the conversion probability is always suppressed by $\varepsilon^2$ and vanishes in the massless dark photon limit, as expected since massless dark photons decouple from the SM.  In the zero background field limit, we have $\Delta\rightarrow0$ and our expression agrees with the usual one in the literature \cite{Masso:2006gc,Ahlers:2007rd,Ahlers:2007qf,An:2013yfc}.

We now turn to some comments about the inhomogeneous background field case. 


\subsection{Inhomogeneous External Field}

In an inhomogeneous external field, the conversion probability in the limit of weak mixing can be computed by the equivalent of time-dependent perturbation theory in quantum mechanics, leading to the approximation \cite{Raffelt:1987im}
\eqna{
P_{\gamma_a\to\gamma_s}(z)&=\left|\int_0^zdz'\,\varepsilon\Delta_\text{D}\,\text{exp}\left\{i\int_0^{z'}dz''\,[\Delta_\text{D}-\Delta(z'')]\right\}\right|^2\\
&=\varepsilon^2\Delta_\text{D}^2\left|\int_0^zdz'\,\text{exp}\left\{i\int_0^{z'}dz''\,[\Delta_\text{D}-\Delta(z'')]\right\}\right|^2.
}[EqnPinhom]
The conversion probability \eqref{EqnPinhom} is correct as long as the numerical value for $\varepsilon$ is small enough for the approximation to make sense.  Moreover, the conversion probability vanishes in the limit $\Delta_\text{D}\to0$, as expected.  Finally, \eqref{EqnPinhom} simplifies to \eqref{EqnP} in the limit of homogeneous external fields.

As for the conversion probability in the homogeneous case, the conversion probability in the inhomogeneous case \eqref{EqnPinhom} is suppressed by $\varepsilon^2$ and vanishes when $\Delta_\text{D}$ is set to zero due to the massless dark photon decoupling.


\section{Conclusion}\label{SConclusion}

The mixing of pseudoscalars with the photon in background electromagnetic fields has led to a vast ecosystem of experimental searches for new physics.  The dark photon version of this story is usually not considered due to the Landau-Yang theorem.  However, due to SM particles coupling to a mixed state of photon and dark photon, conversion does occur.

Our focus in this paper has been to lay the theoretical groundwork for studying the mixing of photons and dark photons in strong background electromagnetic fields.  The relevant diagram is a four-photon vertex obtained by integrating out the electron box diagram in non-linear QED.  We started from the Schwinger Lagrangian and derived the equations of motion for the dark photon and the visible photon.  We then provided expressions for the perpendicular and parallel refractive indices, including plasma effects, as well as the probability of photon-dark photon conversions.

There are several future directions to be pursued.  Firstly, our results should be applied to the extreme environments near the surface of magnetars.  The magnetic field approaches and in some cases exceeds the quantum critical value, although from the homogeneous case we do not expect appreciable photon-dark photon conversion to occur.  Nevertheless, at resonance photon emission from magnetars should undergo possibly measurable attenuation due to this conversion, and this in turn should constrain the photon-dark photon coupling $\varepsilon$.  A detailed treatment would require the probability of conversion in a dipolar magnetic field, and is left for the future.

Secondly, our results should be applied to the extreme environments in the upcoming 10 PW optical laser systems.  These setups target the investigation of non-linear QED as a fundamental physics goal.  To this, one can add the photon-dark photon mixing scenarios as well.  For these lasers, the intensities are somewhat lower than the quantum critical value.  Nevertheless, the effect of dark photons on the dichroism and birefringence of the vacuum in these environments may be interesting.

Thirdly, we note that our results rely on the Schwinger Lagrangian, valid at photon energies below the electron mass.  Thus, our formalism is applicable for photon-dark photon conversions only up to the hard X-ray spectrum.  It would be interesting to extend our analysis to even higher photon energies.


\ack{
The authors would like to thank Andrea Caputo, Hongwan Liu, Siddharth Mishra-Sharma, Maxim Pospelov and Josh Ruderman for pointing out a mistake in the inhomogeneous external field case.  The work of JFF is supported by NSERC and FRQNT.  KS is supported by the U.~S.~Department of Energy grant DE-SC0009956.
}


\bibliography{PhotonDarkPhotonConversion}


\end{document}